\documentclass[
  prc
 ,superscriptaddress
 ,twocolumn
 ,showpacs
 ,preprintnumbers
]{revtex4-1}
\usepackage{graphicx}
\usepackage{amsmath,amssymb}
\usepackage{newtxtext,newtxmath}
\usepackage{bm}
\usepackage{braket}

\usepackage{hyperref}
\hypersetup{pdfpagemode={UseOutlines},
bookmarksopen=true,
bookmarksopenlevel=0,
hypertexnames=false,
colorlinks=true,citecolor=blue,linkcolor=blue,urlcolor=blue,allcolors=blue,pdfstartview={FitV},
unicode,
breaklinks=true,
}

\begin{document}
\preprint{NITEP 22}
\title{
  Quantitative description of the $^{20}$Ne($p$,$p\alpha$)$^{16}$O reaction as a means of probing the surface $\alpha$ amplitude 
}

\author{Kazuki~Yoshida}
\email[]{yoshida.kazuki@jaea.go.jp}
\affiliation{Advanced Science Research Center, Japan Atomic Energy Agency,
Tokai, Ibaraki 319-1195, Japan}
\affiliation{Research Center for Nuclear Physics (RCNP), Osaka
University, Ibaraki 567-0047, Japan}

\author{Yohei~Chiba}
\affiliation{Department of Physics, Osaka City University, Osaka 558-8585, Japan}
\affiliation{Nambu Yoichiro Institute of Theoretical and Experimental Physics (NITEP), Osaka City University, Osaka 558-8585, Japan}
\affiliation{Research Center for Nuclear Physics (RCNP), Osaka
University, Ibaraki 567-0047, Japan}

\author{Masaaki~Kimura}
\affiliation{Department of Physics, Hokkaido University, Sapporo 060-0810, Japan}
\affiliation{Nuclear Reaction Data Centre, Hokkaido University, Sapporo 060-0810, Japan}
\affiliation{Research Center for Nuclear Physics (RCNP), Osaka
University, Ibaraki 567-0047, Japan}

\author{Yasutaka~Taniguchi}
\affiliation{Department of Information Engineering, National Institute of Technology (KOSEN), Kagawa
College, Mitoyo, Kagawa 769-1192, Japan}
\affiliation{Research Center for Nuclear Physics (RCNP), Osaka
University, Ibaraki 567-0047, Japan}

\author{Yoshiko~Kanada-En'yo}
\affiliation{Department of Physics, Kyoto University, Kyoto 606-8502, Japan}
\affiliation{Research Center for Nuclear Physics (RCNP), Osaka
University, Ibaraki 567-0047, Japan}

\author{Kazuyuki~Ogata}
\affiliation{Research Center for Nuclear Physics (RCNP), Osaka
University, Ibaraki 567-0047, Japan}
\affiliation{Department of Physics, Osaka City University, Osaka 558-8585, Japan}
\affiliation{Nambu Yoichiro Institute of Theoretical and Experimental Physics (NITEP), Osaka City University, Osaka 558-8585, Japan}
\date{\today}

\begin{abstract}
\begin{description}

\item[Background]
  The proton-induced $\alpha$ knockout reaction has been utilized for decades to investigate
  the $\alpha$ cluster formation in the ground state of nucleus.
  However, even today, the theoretical description of the reaction is not precise enough for
  the quantitative study.
  For example,
  the $\alpha$ spectroscopic factors reduced from $\alpha$ knockout 
  experiments with reaction analyses using phenomenological $\alpha$ cluster wave functions disagree 
  with those given by a structure theory.
  In some cases they also scatter depending on the kinematical condition of the experiment.
  This suggests that the theoretical description of the $\alpha$ knockout reaction is insufficient 
  from a quantitative viewpoint.
\item[Purpose]
  We show that the distorted wave impulse approximation can describe 
  $^{20}$Ne($p$,$p\alpha$)$^{16}$O reaction quantitatively if reliable inputs are used;
  the optical potential, the $p$-$\alpha$ cross section, and the $\alpha$ cluster wave function.
  We also investigate the relationship between the $\alpha$ cluster wave function and 
  the $\alpha$ knockout cross section.
\item[Method]
  The $^{20}$Ne($p$,$p\alpha$)$^{16}$O reaction
  is described by the distorted wave impulse approximation.
  An input of the calculation, the $\alpha$-$^{16}$O cluster wave function, is obtained by  
  the antisymmetrized molecular dynamics and the Laplace expansion method.
\item[Results]
  In contrast to the previous work,
  the $^{20}$Ne($p$,$p\alpha$)$^{16}$O data at 101.5~MeV is successfully reproduced
  by the present framework without any free adjustable parameters.
  It is also found that 
  the knockout cross section is sensitive to the surface region of the cluster
  wave function because of the peripherality of the reaction.
\item[Conclusions]
  Using a reliable $\alpha$ cluster wave function, 
  $p$-$\alpha$ cross section, and distorting 
  potentials, it is found that
  the $^{20}$Ne($p$,$p\alpha$)$^{16}$O cross section is 
  quantitatively reproduced by the present framework.
  This success demonstrates that the proton-induced $\alpha$ knockout reaction 
  is a quantitative probe for the $\alpha$ clustering.
\end{description}
\end{abstract}


\maketitle

\section{Introduction}
As it is schematically illustrated in the Ikeda diagram~\cite{Ikeda68},
the $\alpha$ particle is expected to emerge as a subunit in nuclear systems,
in the light mass region in particular,
reflecting the fact that the $\alpha$ particle is a tightly bound system of
four nucleons with the large first excitation energy compared to neighboring nuclei.
Various cluster models have been developed so far,
and in recent decades microscopic cluster models based on the nucleon
degrees of freedom with fermionic quantum statistics
are available, as reviewed in a recently published article~\cite{Freer18}.
Among them,
the antisymmetrized molecular dynamics (AMD) has been applied to many systems and
succeeded in describing cluster structures
~\cite{Enyo95,Enyo98,Enyo01,Kimura04,Enyo12,Kimura16,Chiba16,Enyo16_1,Enyo16_2,Chiba17}.
One of the questions remaining today is 
to what extent $\alpha$ cluster state exists in the nuclear ground states that lie
far below the $\alpha$ threshold.

From a nuclear reaction point of view, the proton-induced $\alpha$ knockout reaction,
($p$,$p\alpha$), is considered to be a good probe for the $\alpha$ cluster state in the ground state of a
target nucleus.
Much effort has been made on the ($p$,$p\alpha$) reaction
studies~\cite{Roos77,Nadasen80,Carey84,Wang85,Nadasen89,Yoshimura98,Neveling08,Mabiala09,Yoshida16,Lyu18,Yoshida18},
but even today quantitative understanding of the ($p$,$p\alpha$) cross section
and its relation with the $\alpha$ cluster wave function
in the ground state of the target nucleus have not yet fully established.
For example, 
in Ref.\cite{Carey84} 
it is reported that the $\alpha$ spectroscopic factor ($S_\alpha$) of the $\alpha$-$^{16}$O cluster
state in the ground state of $^{20}$Ne deduced from the $^{20}$Ne($p$,$p\alpha$)$^{16}$O
experiment is 2.4--3.0 times as large as those given by structure theories.
It is also reported in Ref.~\cite{Mabiala09} that
$S_\alpha$ of $^{12}$C from the
$^{12}$C($p$,$p\alpha$)$^{8}$Be reaction at 100~MeV has large uncertainty of
$S_\alpha=0.19$--$1.68$ depending on the kinematics of the experiment,
while $S_\alpha$ should be determined purely from the structure of $^{12}$C
and should not depend on the kinematical condition of the reaction.
Those uncertainties may arise from the ambiguities in the reaction analyses,
namely, the optical potentials, the $\alpha$ cluster wave function, and
the $p$-$\alpha$ differential cross section and its energy and angular dependence. 

Considering the above-mentioned situation, 
a qualitative description of the ($p$,$p\alpha$) reaction is still challenging.
In the present study, we aim to reproduce 
the $^{20}$Ne($p$,$p\alpha$)$^{16}$O cross section data~\cite{Carey84} 
within the distorted wave impulse approximation (DWIA) framework~\cite{Chant83,Wakasa17}.
Since the target and the residue of this reaction are typical stable nuclei and
the distorting potentials of them are well known,
the description of the reaction will be free from ambiguities
of the optical potential.
In addition, in the experiment the emission angles of the $p$ and $\alpha$ are fixed
and the corresponding $p$-$\alpha$ binary scattering angle is very limited.
This may also help to reduce the ambiguity arising from the angular dependence 
of the $p$-$\alpha$ cross section
required in the DWIA calculation.
As for the $\alpha$ cluster wave function,
the $\alpha+^{16}$O reduced width amplitude (RWA) of the $^{20}$Ne ground state
obtained by the AMD framework~\cite{Chiba17} is adopted
as an input of the DWIA calculation.

In Sec.~\ref{sectheory},
the theoretical description of AMD and DWIA for the $\alpha+^{16}$O cluster
state and the $^{20}$Ne($p$,$p\alpha$)$^{16}$O reaction, respectively, are introduced.
In Sec.~\ref{secresult}, numerical input for the calculations
is explained, and a comparison between the obtained $^{20}$Ne($p$,$p\alpha$)$^{16}$O
cross section and the experimental data are shown.
The probed region of the reaction is also investigated
by applying several Brink-Bloch 
wave functions~\cite{Brink66} of the $\alpha$-$^{16}$O cluster state
to the reaction calculation.
Finally, a summary is given in Sec.~\ref{secsummary}.

\section{Theoretical framework}
\label{sectheory}
\subsection{Reduced width amplitude}
The RWA is the probability amplitude to find the clusters at intercluster distance $R$, and is defined as the
overlap between the ground state wave function of $^{20}{\rm Ne}$ and the reference wave function 
for the $\alpha$+$^{16}{\rm O}$ clustering,
\begin{align}
y(R) = \sqrt{\frac{20!}{16! 4! 4\pi}}
 \Braket{\frac{\delta(r-R)}{r^2}\Phi_{\alpha}\Phi_{^{16}\mathrm{O}}|
  \Psi_{^{20}{\rm Ne}}}.\label{eq:rwa1}
\end{align}
The integral of the RWA called $\alpha$ spectroscopic factor $S_\alpha$ is often used as a measure of
the clustering,
\begin{align}
 S_\alpha = \int_0^\infty R^2 dR |y(R)|^2.
\end{align}
In the definition of the RWA given by Eq.~(\ref{eq:rwa1}), the bra vector represents the reference
cluster state, in which the $\alpha$ and $^{16}{\rm O}$ clusters are coupled to angular momentum
zero with the inter-cluster distance $R$. The ground state wave functions of the $\alpha$ and
$^{16}{\rm O}$ clusters ($\Phi_{\alpha}$ and $\Phi_{^{16}{\rm O}}$) are the harmonic oscillator
wave functions of double closed shell configurations. The ket state $\Psi_{^{20}{\rm Ne}}$
is the ground state wave function of $^{20}{\rm Ne}$ calculated by AMD. 
The AMD wave function is a superposition of parity and angular momentum projected Slater
determinants, and it was shown that the known properties of the $^{20}$Ne ground and excited
states such as the radius, energies, electromagnetic transitions, and $\alpha$ decay widths are
reasonably described. 
Once the ground state wave function is obtained, Eq.~(\ref{eq:rwa1}) is
calculated by using the Laplace expansion method \cite{Chiba17}. For more details of the
calculation, readers are directed to Refs.~\cite{Kimura04,Chiba16}.
It should be noted that other cluster models yield similar RWA and 
reasonably reproduce the observed decay widths of the excited states
\cite{Nemoto72,Matsuse75,Fujiwara79,Enyo14,Kimura04}.
The use of AMD in the present study is aimed to extend our framework 
to the investigation of the $\alpha$ clustering
in unstable nuclei in the future.

\subsection{Distorted wave impulse approximation}
We employ the DWIA framework
in the present study.
Details of DWIA for the description of the knockout reaction can
be found in a recent review paper~\cite{Wakasa17}.
The incident and emitted protons are labeled as particles 0 and 1, respectively.
The wave number and its solid angle, the total and kinetic energies
of particle $i$ ($=0, 1, \alpha$) are represented by
$\bm{K}_i$, $\Omega_i$, $E_i$, and $T_i$, respectively.
Quantities with (without) the superscript L are evaluated in the
laboratory (center-of-mass) frame.

The triple differential cross section (TDX) of the $A$($p$,$p\alpha$)$B$ reaction within
the so-called factorized form of the DWIA framework is given by
\begin{align}
  \frac
  {d^3\sigma}
  {dE_1^\mathrm{L} d\Omega_1^\mathrm{L} d\Omega_2^\mathrm{L}}
  &=
  F_{\mathrm{kin}}C_0
  \frac
  {d\sigma_{p\alpha}}
  {d\Omega_{p\alpha}}
  (\theta_{p\alpha},T_{p\alpha})
  \left|
  \bar{T}_{\bm{K}_i}
  \right|^2.
  \label{eq_tdx}
\end{align}
It is essentially a product of the 
absolute square of the reduced transition matrix $\bar{T}_{\bm{K}_i}$
and the $p$-$\alpha$ two-body differential cross section $d\sigma_{p\alpha}/d\Omega_{p\alpha}$
at a given scattering angle (energy) $\theta_{p\alpha}$ ($T_{p\alpha}$).
To relate an off-the-energy-shell $p$-$\alpha$ scattering in the ($p,p\alpha$) three-body kinematics
to an on-shell observable,
the final-state prescription of the on-shell approximation
is adopted; 
$T_{p\alpha}$ is determined by the
$p$-$\alpha$ relative momentum in the final state.
The kinematical factor (or also referred to as the phase space factor) $F_{\mathrm{kin}}$
and a constant $C_0$ are defined by 
\begin{align}
  F_{\mathrm{kin}}
  &\equiv
  J_{\mathrm{L}}
  \frac
  {K_1 K_\alpha E_1 E_\alpha}
  {(\hbar c)^4}
  \left[
    1+\frac{E_\alpha}{E_\mathrm{B}}
    +\frac{E_\alpha}{E_\mathrm{B}}\frac{\bm{K}_1\cdot\bm{K}_\alpha}{K_\alpha^2}
  \right], \\
  C_0
  &\equiv
  \frac{E_0}{(\hbar c)^2 K_0}
  \frac{\hbar ^4}{(2\pi)^3\mu_{p\alpha}^2},
\end{align}
where $\mu_{p\alpha}$ is the reduced mass of the $p$-$\alpha$ binary system
and $J_{\mathrm{L}}$ is the Jacobian from the center-of-mass frame to the L frame.
The reduced transition matrix is given by 
\begin{align}
  \bar{T}_{\bm{K}_i}
  &\equiv
  \int d \bm{R}\,
  F_{\bm{K}_i}(\bm{R})y(R)Y_{00}(\hat{\bm{R}}), 
  \label{eq_Tbar}
  \\
  F_{\bm{K}_i}(\bm{R})
  &\equiv
  \chi_{1,\bm{K}_1}^{*(-)}(\bm{R})\,
  \chi_{\alpha,\bm{K}_\alpha}^{*(-)}(\bm{R})\,
  \chi_{0,\bm{K}_0}^{ (+)}(\bm{R})\,
  e^{-i\bm{K}_0\cdot{\bm{R}A_\alpha/A}},
  \label{eq_fki}
\end{align}
where $A_\alpha=4$, $A=20$, and $\chi_{i,{\bm{K}_i}}$ is a distorted wave between particle $i$ and $A$ when $i=0$,
otherwise between $i$ and $B$.
The outgoing and incoming boundary conditions of the scattering waves
are specified by the superscripts $(+)$ and $(-)$, respectively.
Note that the $S_\alpha$ does not appear explicitly in Eqs.~(\ref{eq_tdx})--(\ref{eq_fki})
because it is already taken into account in the RWA.

\section{Results and discussion}
\label{secresult}
\subsection{Numerical inputs}
\begin{figure}[htbp]
\centering
\includegraphics[width=0.43\textwidth]{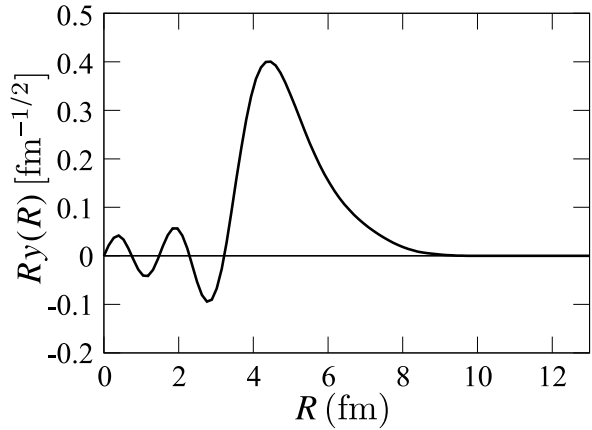}
\caption{
  $\alpha+^{16}$O RWA of the $0_1^+$ state taken from Fig.~2 of Ref.~\cite{Chiba17}.
}
\label{fig_rwa}
\end{figure}

The RWA is calculated by AMD and the Laplace
expansion method~\cite{Chiba17}. The Hamiltonian used in the AMD calculation is given by 
\begin{align}
  \hat{H}
  &=
  \sum_{i} \hat{t}_i - \hat{t}_{\mathrm{c.m.}}
  +\sum_{i<j} \hat{v}_{NN}
  +\sum_{i<j} \hat{v}_{\mathrm{Coul}},
\end{align}
where $\hat{t}_i, \hat{t}_{\mathrm{c.m.}}, \hat{v}_{NN}$, and $\hat{v}_{\mathrm{Coul}}$
being
the nucleon and the center-of-mass kinetic energies,
effective nucleon-nucleon interaction, and the Coulomb interaction, respectively.
As for $\hat{v}_{\mathrm{NN}}$, the Gogny D1S interaction~\cite{Berger91} is adopted.
The obtained RWA \cite{Chiba17} (AMD-RWA) is shown in Fig.~\ref{fig_rwa}, and its $\alpha$ spectroscopic
factor is $S_\alpha=0.26$. 

\begin{figure}[htbp]
\centering
\includegraphics[width=0.43\textwidth]{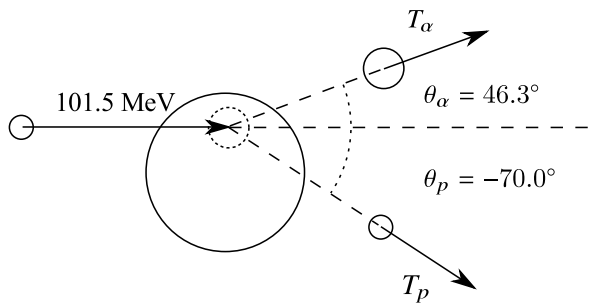}
\caption{
  Kinematical condition of the $^{20}$Ne($p$,$p\alpha$)$^{16}$O reaction~\cite{Carey84}.
  }
\label{fig_kinematics}
\end{figure}

The kinematical condition of the $^{20}$Ne($p$,$p\alpha$)$^{16}$O reaction
is shown in Fig.~\ref{fig_kinematics}.
In the experiment~\cite{Carey84} the so-called energy sharing distribution was measured; 
the emitted angle of $p$ ($\alpha$) is fixed at $-70^\circ$ ($46.3^\circ$),
and the proton emission energy $T_p$ is varied in the range of $40$--$75$~MeV.
By the energy conservation,
$T_\alpha$ is ranging from $55$ to $21$~MeV accordingly.
The incident energy is set to 101.5~MeV and all the
scattering particles are kept in the same plane.
With such a kinematical setup with no recoil, i.e.,
the reaction residue being at rest in the final state, is achieved at around
$T_p=67$~MeV ($T_\alpha=30$~MeV). 
The TDX has a peak at this condition
reflecting the fact that the struck $\alpha$ is bound by an $s$ orbit.
For the optical potential of the incident and emitted protons,
the EDAD1 optical potential of the Dirac phenomenology~\cite{Hama90,Cooper93} is adopted.
The highly reliable optical model parametrization
by F.~Michel \textit{et al}.~\cite{Michel83} is
adopted for the $\alpha$-$^{16}$O scattering in the final state.
As shown in Fig.~1 of Ref.~\cite{Michel83},
this parametrization of the $\alpha$-$^{16}$O optical potential
provides excellent agreement with the $\alpha$-$^{16}$O elastic scattering 
data up to very backward angles, ranging 30--150~MeV $\alpha$ incident energies.

\begin{figure}[htbp]
\centering
\includegraphics[width=0.43\textwidth]{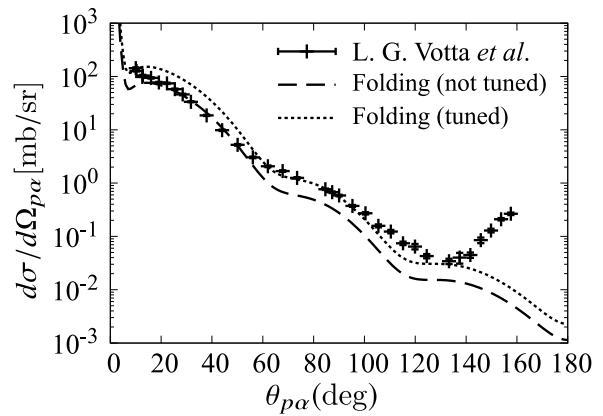}
\caption{
  $p$-$\alpha$ differential cross section at 85~MeV obtained by the
  folding model~\cite{Toyokawa13} with the Melbourne \textit{g}-matrix interaction~\cite{Amos00}
  (dashed line).
  The tuned result at around $\theta_{p\alpha}=85^\circ$ is shown in dotted line.
  The experimental data are taken from Ref.~\cite{Votta74}.
}
\label{fig_p-a}
\end{figure}

Since $\theta_p$ and $\theta_\alpha$ are fixed
in the present ($p$,$p\alpha$) kinematics,
the required $p$-$\alpha$ differential cross section lies
in the very limited range of
$\theta_{p\alpha}=84^\circ$--$86^\circ$ in the $p$-$\alpha$ center-of-mass frame
and $E_{p\alpha}=75$--$100$~MeV in the $p$-$\alpha$ laboratory frame.
Its energy and angular dependence are
obtained by the microscopic single-folding model~\cite{Toyokawa13} with a
phenomenological $\alpha$ density and the Melbourne nucleon-nucleon \textit{g}-matrix
interaction~\cite{Amos00}.
As shown in the dashed line in Fig.~\ref{fig_p-a}, the calculated $p$-$\alpha$
differential cross section at 85~MeV deviates from 
the experimental data~\cite{Votta74} 
by about a factor of 2 at around $\theta_{p\alpha}=84^\circ$--$86^{\circ}$.
In the following analysis
the cross section is tuned by a factor of 2.0 (dotted line)
to correctly reproduce the $p$-$\alpha$ cross section data.
It should be noted that this correction is valid
thanks to the limited range of required $\theta_{p\alpha}$,
which is the outcome of the kinematical condition 
in which $\theta_p$ and $\theta_\alpha$ are fixed
in the $^{20}$Ne($p$,$p\alpha$)$^{16}$O experiment.

\subsection{$\alpha$ cluster wave function and $^{20}$Ne($p$,$p\alpha$)$^{16}$O cross section}
\label{subsec_tdx}
\begin{figure}[htbp]
\centering
\includegraphics[width=0.43\textwidth]{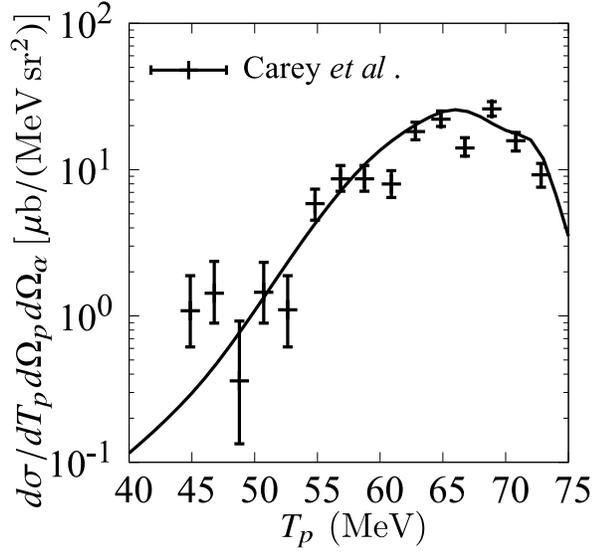}
\caption{
  The comparison between the calculated energy sharing distribution
  with the AMD-RWA (solid line)
  and the experimental data taken from Ref.~\cite{Carey84}.
}
\label{fig_carey}
\end{figure}

In Fig.~\ref{fig_carey} the energy sharing distribution calculated by DWIA with the AMD-RWA is shown.
The present framework well reproduces both the height and distribution 
of the experimental data~\cite{Carey84}
without any additional adjustment.
Thus, the original problem, the inconsistency between the 
reaction analysis and the nuclear structure calculations, is resolved
in the present approach.
It should be stressed that
the quantitative reproduction of the cross section data is guaranteed by the
sophisticated RWA in $^{20}$Ne, by the proper $p$-$\alpha$ cross section, 
and by the use of an appropriate $\alpha$-$^{16}$O optical potential.
In particular, $\alpha$-$^{16}$O optical potential suggested in Ref.~\cite{Michel83}
was essential for this success of quantitative description.
This result indicates that once the proper ingredients mentioned above  
are adopted, the ($p$,$p\alpha$) reaction can be a quantitative probe for the
$\alpha$ amplitude in the ground state of target nuclei.

\begin{figure}[htbp]
\centering
\includegraphics[width=0.43\textwidth]{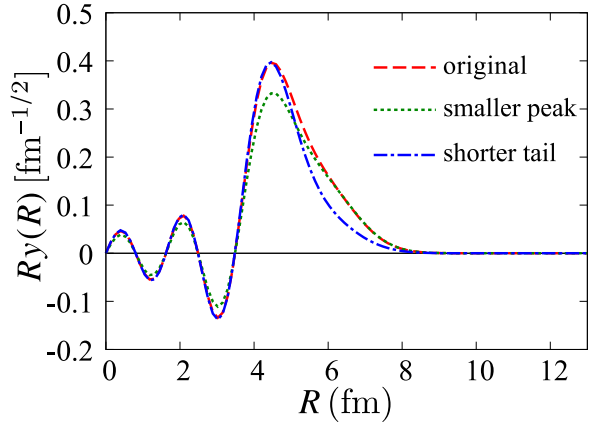}
\caption{
  RWAs constructed by the superposition of two Brink-Bloch wave functions
  of the $\alpha$-$^{16}$O cluster.
  See text for details.
}
\label{fig_rwa_diff}
\end{figure}

Next, we investigate the relationship between the $\alpha$ cluster wave function and 
the $\alpha$ knockout cross section.
Three different types of cluster wave functions shown in Fig.~\ref{fig_rwa_diff} are considered
as artificial input data.
They are constructed by superposing two Brink-Bloch (BB) wave functions~\cite{Brink66}
of the $\alpha$-$^{16}$O cluster.
The inter-cluster distances of the two BB wave functions are fixed at 3.0~fm
and 5.5~fm, whereas the amplitude, $C_1$ and $C_2$, 
of these are varied by hand.
In Table~\ref{table_brink}, we show $C_i$ ($i=1$ or $2$) and 
the resulting $S_\alpha$ of the three RWAs.
\begin{table}[htbp]
\caption{
  Coefficients
  for the BB wave functions.  The resulting $S_\alpha$
  are also shown.
}
  \begin{tabular}{c|ccc}
               & \,original & \,smaller peak & \,shorter tail \\ \hline
    $C_1$      & 0.67       & 0.55           & 0.69      \\
    $C_2$      & 0.17       & 0.17           & 0.09      \\
    $S_\alpha$ & 0.24       & 0.18           & 0.21      \\
  \end{tabular}
\label{table_brink}
\end{table}
As shown in Fig.~\ref{fig_rwa_diff},
$C_i$ of one labeled ``original'' (dashed) is determined so as to reproduce the largest peak and
the tail behavior of the AMD-RWA as much as possible,
while for ``smaller peak'' (dotted) and ``shorter tail'' (dot-dashed),
they are tuned to reduce the peak and tail region of the original RWA, respectively.

To investigate 
how these RWAs contribute to the TDX at the peak of the energy sharing distribution,
it is useful to consider the transition matrix density (TMD) defined by~\cite{Wakasa17}
\begin{align}
  \delta^{\mathrm{Tr}}(R)
  &\equiv
  \bar{T}^{*}_{\bm{K}_i}
  \int d\Omega\,
  R^2 F_{\bm{K}_i}(\bm{R})y(R)Y_{00}(\hat{\bm{R}}).
  \label{eq_tmd}
\end{align}
From Eq.~({\ref{eq_Tbar})
it has the following properties:
\begin{align}
  \int dR\,
  \mathrm{Re}[\delta^{\mathrm{Tr}}(R)]
  &=
  |\bar{T}_{\bm{K}_i}|^2,\\
  \int dR\,
  \mathrm{Im}[\delta^{\mathrm{Tr}}(R)]
  &=
  0.
\end{align}
Thus, one may regard $\mathrm{Re}[\delta^{\mathrm{Tr}}(R)]$ as a ``radial distribution of
the cross section''
and its integrated value is proportional to the TDX.
In Fig.~\ref{fig_tmd_brink} $\mathrm{Re}[\delta^{\mathrm{Tr}}(R)]$ 
at the recoilless condition ($T_p = 67$~MeV) is shown.
\begin{figure}[htbp]
\centering
\includegraphics[width=0.43\textwidth]{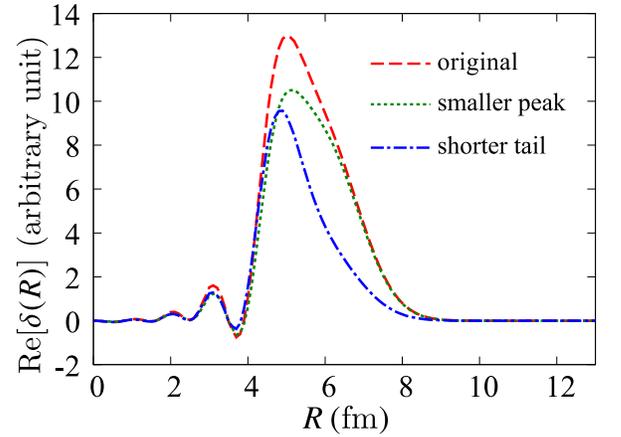}
\caption{
  Real part of the TMD at the recoilless condition: $T_p = 67$~MeV.
  The dashed, dotted, and dot-dashed lines are the results with the
  RWAs shown by dashed, dotted, and dot-dashed lines in Fig.~\ref{fig_rwa_diff}, respectively.
}
\label{fig_tmd_brink}
\end{figure}
It is clearly seen that the internal region of the RWAs are suppressed by the absorption effect
and the surface region contributes to the ($p$,$p\alpha$)
cross section.
In Fig.~\ref{fig_tdx_brink} 
the energy sharing distributions with 
the RWAs of Fig.~\ref{fig_rwa_diff} are shown.
The peak height of the energy sharing distribution
is reduced significantly in the shorter tail case (dot-dashed), 
while it is similar to the
original one in the smaller peak case (dotted),
because of the peripherality of the reaction.
\begin{figure}[htbp]
\centering
\includegraphics[width=0.43\textwidth]{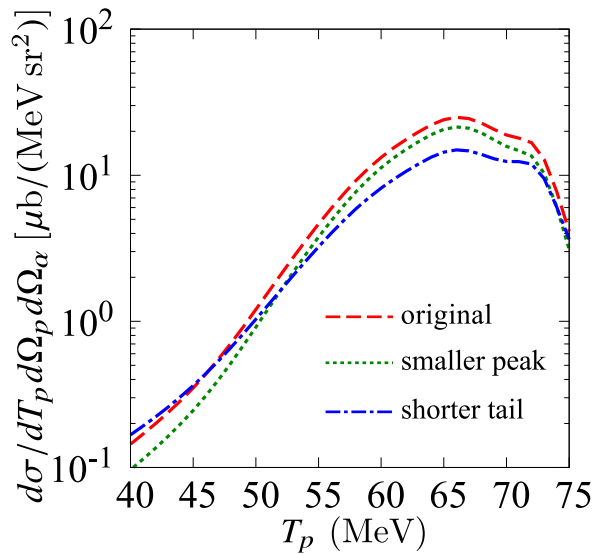}
\caption{
  Energy sharing distributions with the RWAs shown in Fig.~\ref{fig_rwa_diff}.
}
\label{fig_tdx_brink}
\end{figure}
Another response of the difference in the RWAs is the width of the
energy sharing distribution.
Since the RWA with smaller peak (shorter tail) has narrower (wider) momentum distribution, 
its energy sharing distribution has a narrower (wider) width accordingly.
From these results it is shown that the ($p$,$p\alpha$) reaction is a
quantitative probe for the $\alpha$ cluster state, 
putting emphasis on the $\alpha$ amplitude around the nuclear surface.

\section{Summary}
\label{secsummary}
The $^{20}$Ne($p$,$p\alpha$)$^{16}$O reaction at 101.5~MeV was investigated
within the DWIA framework.
AMD was adopted to describe the
$\alpha$-$^{16}$O cluster state of $^{20}$Ne and its RWA
was obtained by the Laplace expansion method.
It was found that by using reliable
$p$-$\alpha$ differential cross section, distorting potentials
as well as the AMD-RWA, the present DWIA calculation quantitatively 
reproduces the experimental data without any additional correction or scaling
in describing the ($p$,$p\alpha$) cross section.
Thus, the inconsistency between the $^{20}$Ne($p$,$p\alpha$)$^{16}$O
reaction analysis and the nuclear structure calculations is clearly resolved
in the present approach.

By the analyses using the BB wave functions with
different spatial distributions and corresponding $\alpha$ spectroscopic 
factors,
it was shown that the surface region of the target is selectively probed
by the $^{20}$Ne($p$,$p\alpha$)$^{16}$O reaction.
Therefore 
the ($p$,$p\alpha$) reaction is a good probe for the surface amplitude of the
cluster wave function, which should be directly related to the 
$\alpha$ clustering of interest.
One may also deduce the momentum distribution of the $\alpha$ cluster wave function
from the width of the energy sharing distribution.
For the momentum distribution of the $\alpha$ cluster to be extracted,
the ($p$,$p\alpha$) cross section should be measured very precisely down to the tail region.

\section*{ACKNOWLEDGMENTS}
This work was
supported in part by Grants-in-Aid of the Japan Society for
the Promotion of Science (Grants No.~JP18K03617,
JP16K05352, JP19K03859, and JP16K05339),
the Hattori Hokokai Foundation Grant-in-Aid for Technological and
Engineering Research,
and by the grant for the RCNP joint research project.

\bibliography{ref}

\end{document}